\documentclass[submission,copyright,
creativecommons]{eptcs}
\providecommand{\event}{ICE 2021} %
\usepackage{underscore}           %

\usepackage{latex/packages}
\usepackage{latex/styles-macros}

\title{Processes, Systems \& Tests: Defining Contextual Equivalences}
\author{Clément Aubert%
	\institute{School of Computer and Cyber Sciences, Augusta University, Georgia, USA }
	\email{caubert@augusta.edu}
	\and
	Daniele Varacca
	\institute{LACL, Université Paris-Est Créteil, France}
	\email{daniele.varacca@u-pec.fr}
}
\def\titlerunning{Processes, Systems \& Tests}
\def\authorrunning{C. Aubert \& D. Varacca}

\begin{document}
\maketitle

\begin{abstract}
	In this position paper, we would like to offer and defend a template to study equivalences between programs---in the particular framework of process algebras for concurrent computation.
	We believe that our layered model of development will clarify the distinction that is too often left implicit between the tasks and duties of the programmer and of the tester.
	It will also enlighten pre-existing issues that have been running across process algebras such as the calculus of communicating systems, the \(\pi\)-calculus---also in its distributed version---or mobile ambients.
	Our distinction starts by subdividing the notion of process in three conceptually separated entities, that we call \emph{Processes}, \emph{Systems} and \emph{Tests}.
	While the role of what can be observed and the subtleties in the definitions of congruences have been intensively studied, the fact that \emph{not every process can be tested}, and that \emph{the tester should have access to a different set of tools than the programmer} is curiously left out, or at least not often formally discussed.
	We argue that this blind spot comes from the under-specification of contexts---environments in which comparisons occur---that play multiple distinct roles but supposedly always \enquote{stay the same}.
\end{abstract}

\section{Introduction}
\label{sec:intro}
In the study of programming languages, contextual equivalences play a central role: to study the behavior of a program, or a process, one needs to observe its interactions with different environments, \eg what outcomes it produces. 
If the program is represented by a term in a given syntax, environments are often represented as contexts surrounding the terms.
But contexts play multiple roles that serve different actors with different purposes.
The programmer uses them to construct larger programs, the user employs them to provide input and obtain an output, and the tester or attacker uses them to debug and compare the program or to try to disrupt its intended behavior.

We believe that representing those different purposes with the same \enquote{monolithic} syntactical notion of context forced numerous authors to repeatedly \enquote{adjust} their definition of context without always acknowledging it.
We also argue that collapsing multiple notions of contexts into one prevented further progress.
In this article, we propose a way of clarifying how to define contextual equivalences, and show that having co-existing notions of equivalences legitimates and explains recurring choices, and supports a rigorous guideline to separate the development of a program from its usage and testing.

Maybe in the mind of most of the experts in the formal study of programming language is our proposal too obvious to discuss.
However, if this is the case, we believe that this \enquote{folklore} remains unwritten, and that since we were not at \emph{that} \enquote{seminar at Columbia in 1976}\footnote{To re-use in our setting Paul Taylor's witty comment~\cite{Taylor2012}.}, we are to remain in darkness.

We believe the interactive and friendly community of ICE to be the ideal place to reach curious and open-minded actors in the field, and to either be proven wrong or---hopefully---impact some researchers.
Non-technical articles can at times have a tremendous impact~\cite{Dijkstra1968}, and even if we do not claim to have Dijkstra's influence or talent, we believe that our precise, documented proposition can shed a new light on past results, and frame current reflections and future developments.
We also believe that ignoring or downplaying the distinctions we stress have repeatedly caused confusions.%

\section{The Flow of Testing}
\label{sec:Java-example}

We begin by illustrating with a simple Java example the three syntactic notions--\emph{process}, \emph{system} and \emph{test}--we will be using.
Imagine giving a user the code \mintinline{java}{while(i < 10){x *= x; i++;}}. %
A user cannot execute or use this \enquote{snippet} unless it is \emph{wrapped} into a method, with adequate header, and possibly variable declaration(s) and \mintinline{Java}{return} statement.
Once the programmer performed this operation, the user can \emph{use} the obtained program, and the tester can \emph{interact} with it further, \eg by calling it from a \mintinline{Java}{main} method.

All in all, a programmer would build on the snippet, then the tester would build an environment to interact with the resulting program, and we could obtain the code below. %
Other situations could arise--\eg, if the snippet was already wrapped--, but we believe this is a fair rendering of \enquote{the life of a snippet}.

\vspace{\baselineskip}
\begin{minipage}[b]{.45\textwidth}
\begin{minted}{java}
|\tikzmark{interact1top}|public class Main{
  |\tikzmark{wrap1top}|public static int foo(int x){
    int i = 0;
    |\tikzmark{code1top}|while(i < 10){
      x *= x;
      i++;
    }|\tikzmark{code1bottom}|
    return x;
  }|\tikzmark{wrap1bottom}|
  public static void main(){
    System.out.print(foo(2));
  }
}|\tikzmark{interact1bottom}|
\end{minted}

\AddToShipoutPictureBG*{
	\begin{tikzpicture}[remember picture, overlay]
	\path[overlay, fill=interact] ([xshift=-2em] interact1top.north) rectangle ([xshift=.4\linewidth] interact1bottom.south) node[above left= -0.1 and 0]{\mintinline{Java}{// Interaction}};
	\path[overlay, fill=wrap] ([xshift=-1em] wrap1top.north) rectangle ([xshift=.3\linewidth] wrap1bottom.south) node[above left= -0.1 and 0]{\mintinline{Java}{// Wrapping}};
	\path[overlay, fill=code] ([xshift=-.5em] code1top.north) rectangle ([xshift=.2\linewidth] code1bottom.south) node[above left= -0.1 and 0]{\mintinline{Java}{// Snippet}};
	\end{tikzpicture}
}
\end{minipage}
\begin{minipage}[b]{.5\textwidth}
In this example, the snippet is what we will call a \emph{process}, the snippet once wrapped is what we will call a \emph{system} and the \enquote{Interaction} part without the system in it, but with additional \enquote{observations}---\ie measures on the execution, terminal output---, is what we will call a \emph{test}.
Our terminology comes from the study of concurrent process algebras, where most of our intuitions and references are located, but let us first make a brief detour to examine how our lens applies to \(\lambda\)-calculus.
\end{minipage}

\section{A Foreword on \texorpdfstring{\(\lambda\)}{Lambda}-Calculus}
\label{sec:lambda}
Theoretical languages often take \(\lambda\)-calculus as a model or a comparison basis. %
It is often said that the \(\lambda\)-calculus is to sequential programs what the  \(\pi\)-calculus is to concurrent programs \cite{Sangiorgi2011a, Varela2013}.
Indeed, pure \(\lambda\)-calculus (\ie without types or additional features like probabilistic sum~\cite{Faggian2019} or quantum capacities~\cite{selinger2009,Tondervan2004}) is a reasonable~\cite{Accattoli2014a}, Turing-complete and elegant language, that requires only a couple of operators%
, one reduction rule %
and one equivalence relation %
to produce a rich and meaningful theory, 
sometimes seen as an idealized target language for functional programming languages.

Since most terms\footnote{Actually, if application, abstraction and variables all count as one, the ratio between normal term and term with redexes is unknown~\cite{Bodini2021}. We imply here \enquote{since \emph{most interesting} terms}, in the sense of terms that represent programs.} do not reduce as they are, to study their behavior, one needs first to make them interact with an environment, represented by a context.
Contexts are generally defined as \textcquote[p.~29, 2.1.18]{Barendregt1984}{term\textins{s} with some holes}, that we prefer to call \emph{slots} and denote  \([\square]\).
Under this apparent simplicity, they should not be manipulated carelessly, as having multiple slots or not being careful when defining what it means to \enquote{fill a slot} can lead to \eg lose confluence~\cite[pp.~40--41, Example 2.2.1]{Mirna2002}, and as those issues persist even in the presence of a typing system~\cite{Hashimoto2001}.
Furthermore, definitions and theorems that use contexts frequently impose some restrictions on the contexts considered, to exclude \eg \((\lambda x. y)[\square]\) that simply \enquote{throw away} the term put in the slot in one step of reduction.
Following the above observations, we conclude that contexts often come in two flavors, depending on the nature of the term considered: %

\begin{description}
	\item[For closed terms] (\ie without free variables), a context is essentially a series of arguments to feed the term.
	      This observation allows to define \eg \emph{solvable terms}~\cite[p.~171, 8.3.1 and p.~416, 16.2.1]{Barendregt1984}.
	\item[For open terms] (\ie with free variables), a context is a \emph{B\"ohm transformation}~\cite[p.~246, 10.3.3]{Barendregt1984}, which is equivalent~\cite[p.~246, 10.3.4]{Barendregt1984} to a series of abstractions followed by a series of applications, and sometimes called \enquote{head context}~\cite[p.~25]{Accattoli2012b}.
\end{description}

Being closed corresponds to being \enquote{wrapped}--ready to use--, and feeding arguments to a term corresponds to interacting with it from a \mintinline{Java}{main} method: the B\"ohm transformation actually encapsulates two operations at once.
In this case, the interaction can observe different aspects: whether the term terminates, whether it grows in size, \etc, but it is generally agreed upon that no additional operator or reduction rule should be used.
Actually, the syntax is restricted when testing, as only application is allowed: the tested term should not be wrapped in additional layers of abstraction if it is already closed.

Adding features to the \(\lambda\)-calculus certainly does not restore the supposed purity or unicity of the concept of context, but actually distances it even further from being simply \enquote{a term with a slot}. For instance, contexts are narrowed down to term context~\cite[p.~1126]{Tondervan2004} and surface context~\cite[pp.~4, 10]{Faggian2019} for respectively quantum and probabilistic \(\lambda\)-calculus, to \enquote{tame} the power of contexts.
In resource sensitive extensions of the \(\lambda\)-calculus, the quest for full abstraction even led to a drastic separation of \(\lambda\)-terms between terms and tests~\cite{bucciarelli2011}, a separation naturally extended to contexts~\cite[p.~73, Figure 2.4]{Breuvart2015}.

This variety happened after the 2000's formal studies of contexts was undertaken~\cite{Mirna2002,Bognar2001,Hashimoto2001}, which led to the observation that treating contexts \textcquote[p.~29]{Bognar2001}{merely as a notation \textelp{} hinders any formal reasoning\textins{, while treating them} as first-class objects \textins{allows} to gain control over variable capturing and, more generally, \enquote{communication} between a context and expressions to be put into its holes}.
It is ironic that \(\lambda\)-calculists took inspiration from a concurrent language to split their syntax in two right at its core~\cite[p.~97]{bucciarelli2011}, or to study formally the \emph{communication} between a context and the term in its slot, while concurrent languages sometimes tried to keep the \enquote{purity} and indistinguishability of their contexts. %
It can be noted that this re-definition of contexts had impacts on other fields, \eg on modal type theory~\cite{Nanevski2008}.

In the case of concurrent calculi like the calculus of communicating systems (CCS) or the \(\pi\)-calculus, interactions with environments are also represented using a notion of context.
But the status of contexts in concurrent calculi is even more unsettling when one notes that, while \enquote{wrapping} contexts are of interest mainly for open terms in lambda calculus, \emph{all} terms need a pertinent notion of context in concurrent systems to be tested and observed.
Indeed, as the notion of \enquote{feeding arguments to a concurrent process} concurs with the idea of wrapping it into a larger process, it seems that the distinction we just made between two kinds of contexts in \(\lambda\)-calculus cannot be ported to concurrent calculi.
Our contribution starts by questioning whenever, indeed, process calculi have treated contexts as a uniform notion independently from the nature of the term or what it was used for.

\section{Contextual Relations}\label{sec:purposes}

Comparing terms is at the core of the study of programming languages, and process algebra is no exception.
Generally, and similarly to what is done in \(\lambda\)-calculus, a comparison is deemed of interest only if it is valid in every possible context\footnote{As a reviewer noted, \enquote{Proving a behavioural equivalence is a congruence has a reading that is somewhat left out of the discussion in the paper. Namely, the behavioural equivalence provides a mathematical notion of behaviour and showing such behaviour is preserved by language contexts attests that the latter are proper functions of behaviour (\ie, if \(C[\cdot]\) is provided with \enquote{behaviour} then the result is still \enquote{behaviour}). In this sense, the congruence result can be viewed as a sanity check on the language constructors.}} an idea formally captured by the notion of (pre-)congruence.
An equivalence relation \(\rel\) is usually said to be a congruence if it is closed by context, \ie if for all \(P\), \(Q\) (open or closed) terms, \((P,  Q) \in \rel\) implies that for all context \(C[\square]\),  \((C[P], C[Q]) \in  \rel \). %
(Sometimes, the additional requirement that terms in the relation needs to be similar up to uniform substitution is added~\cite{Honda1995}, and sometimes~\cite[p.~516, Definition 2]{Parrow2001}, only the closure by substitution---seen as a particular kind of context---is required.)

A notable example of congurence is \emph{barbed congruence}~\cites[Definition 2.1.4]{Madiot2015}[Definition 8]{Milner1992}, which closes by context a reduction-closed relation used to observe \enquote{barbs}--the channel(s) on which a process can emit or receive.
It %
 is often taken to be \emph{the} \textcquote[p.~4]{Madiot2015}{reference behavioural equivalence}, as it observes the interface of processes, \ie on which channels they can interact over the time and in parallel.

But behind this apparent uniformity in the definition of congruences, the definition of contextual relations itself have often been tweaked by altering the definition of context, with no clear explanation nor justification, as we illustrate below. %

\begin{description}[leftmargin=.4em]
	\item[In the calculus of communicating systems,] notions as central as contextual bisimulation~\cite[pp.~223-224, Definition 421]{Amadio2016} and barbed equivalence~\cite[p.~224, Definition 424]{Amadio2016} considers only \emph{static} contexts~\cite[p.~223, Definition 420]{Amadio2016}, which are composed only of parallel composition with arbitrary term and restriction.
	      As the author of those notes puts it himself, \textcquote[p.~227]{Amadio2016}{the rules of the bisimulation game may be hard to justify \textins{and} contextual bisimulation \textelp{} is more natural}. But there is no justification---other than technical, \ie because they \textcquote[p.~223]{Amadio2016}{they persist after a transition}---as to \emph{why} one should consider only some contexts in defining contextual equivalences.

	\item[In the \(\pi\)-calculus,] contexts are defined liberally~\cite[p.~19, Definition 1.2.1]{Sangiorgi2001}, but still exclude contexts like \eg \([\square] + 0\) right from the beginning.
	      Congruences are then defined using this notion of context~\cite[p.~19, Definition 1.2.2]{Sangiorgi2001},
	      and strong barbed congruence is no exception~\cite[p.~59, Definition 2.1.17]{Sangiorgi2001}. Other notions, like strong barbed equivalence~\cite[p.~62, Definition 2.1.20]{Sangiorgi2001}, are shown to be a non-input congruence~\cite[p.~63, Lemma 2.1.24]{Sangiorgi2001}, which is a notion relying on contexts that forbids the slot to occur under an input prefix~\cite[p.~62, Definition 2.1.22]{Sangiorgi2001}.
	      In other words, two notions of contexts and of congruences co-exist generally in \(\pi\)-calculus, but
	      \textcquote[p.~245]{Hennessy2007}{\textins{i}t is difficult to give rational arguments as to why one of these relations is more reasonable than the other.}

	\item[{In the distributed \(\pi\)-calculus,}] contexts are restricted right from the beginning to particular operators~\cite[Definition 2.6]{Hennessy2007}.
	      Then, relations are defined to be contextual if they are preserved by static contexts~\cite[Definition 2.6]{Hennessy2007}, which contains only parallel composition with arbitrary terms and name binding.
These contexts also appear as  \textcquote[p.~375]{Lanese2013}{configuration context} or \enquote{harness} in the ambient calculus~\cite[p.~372]{Gordon2003}.
	      Static operators are deemed \textcquote[p.~37]{Hennessy2007}{sufficient for our purpose} and static contexts only are considered \textcquote[p.~38]{Hennessy2007}{\textins*{t}o keep life simple}, but no further justification is given.

	\item[In the semantic theory for processes,] at least in the foundational theory we would like to discuss below, one difficulty is that the class of formal theories restricted to \textcquote[p.~448]{Honda1995}{reduction contexts} still fall short on providing a satisfactory \enquote{formulation of semantic theories for processes which does not rely on the notion of observables or convergence}.
	      Hence, the authors have to furthermore restrict the class of terms to \emph{insensitive} terms~\cite[p.~450]{Honda1995} to obtain a notion of \emph{generic reduction}~\cite[p.~451]{Honda1995} that allows a satisfactory definition of sound theories~\cite[p.~452]{Honda1995}.
	      Insensitive terms are essentially the collection of terms that do not interact with contexts~\cite[p.~451, Proposition~3.15]{Honda1995}, an analogue to \(\lambda\)-calculuus' genericity Lemma~\cite[p.~374, Proposition~14.3.24]{Barendregt1984}.
	      Here, contexts are restricted by duality: insensitive terms are terms that will \emph{not} interact with the context in which they are placed, and that need to be equated by sound theories.

	\item[Across calculi,] a notion of \enquote{closing context}---that emerged from \(\lambda\)-calculus~\cite[p.~85]{Amadio2016}, and matches the \enquote{wrapping} of a snippet---can be found in \eg typed versions of the \(\pi\)-calculus~\cite[p.~479]{Sangiorgi2001}, in mobile ambient~\cite[p.~134]{Varela2013}, in the applied \(\pi\)-calculus~\cite[p.~7]{Abadi2018}, and in the fusion calculus~\cite[p.~6]{Merro1998}. %
	Also known as \enquote{completing context}~\cite[p.~466]{Sangiorgi1999}, those contexts are parametric in a term, the idea being that such a context will \enquote{close} the term under study, making it amenable to tests and comparisons.
\end{description}

Let us try to extract some general principles from this short survey.
It seems that contexts are
\begin{enumerate*}
	\item \emph{in appearance} given access to the same operators than terms,
	\item sometimes deemed to be \enquote{un-reasonable}, without always a clear justification,
	\item shrunken by need, to bypass some of the difficulties they raise, or to preserve some notions,
	\item sometimes picked by the term itself---typically because the same \enquote{wrapping} cannot be applied to all processes.
\end{enumerate*}
Additionally, in all those cases, contexts are given access to a subset of operators, or restricted to contexts with particular behavior, \emph{but never extended}. %
If we consider that contexts are the main tool to test the equivalence of processes, then why should the testers--or the attacker--always have access to \emph{fewer} tools than the programmer?
What reason is there not to \emph{extend} the set of tools, of contexts, or simply take it to be orthogonal?
The method we sketch below allows and actually encourages such nuances, would justify and acknowledge the restrictions we just discussed instead of adding them \emph{passing-by}, and actually corresponds to common usage. %
\section{Processes, Systems and Tests}
\label{sec:pst}

As in the \(\lambda\)-calculus, most concurrent calculi make a distinction between open and closed terms.
For instance, the distributed \(\pi\)-calculus~\cite{Hennessy2007} implements a distinction between closed terms (called processes~\cite[p.~14]{Hennessy2007}) and open terms, based on binding operators (input and recursion).

Most of the time, and since the origin of the calculus of communicating systems, the theory starts by considering only programs---\textcquote[p.~73]{Milner1980}{closed behaviour expression\textins{s}, \ie ones with no free variable}---when comparing terms, as---exactly like in \(\lambda\)-calculus---they correspond to self-sufficient, well-rounded programs: it is generally agreed upon that open terms should not be released \enquote{into the wild}, as they are not able to remain in control of their internal variables, and can be subject to undesirable or uncontrolled interferences.
Additionally, closed terms are also the only ones to have a \emph{reduction semantics}, which means that they can evolve without interacting with the environment--this would corresponds, in Java, to being wrapped, \ie inserted into a proper header and ready to be used or tested.

However, in concurrent calculi, the central notions of binders and of variables have been changing, and still seem today sometimes \enquote{up in the air}. For instance, in the original CCS, restriction was not a binder~\cite[p.~68]{Milner1980}, and by \textcquote[p.~16]{Milner1989}{refusing to admit channels as entities distinct from agents} and defining two different notions of scopes~\cite[p.~18]{Milner1989}, everything was set-up to produce a long and recurring confusion as to what a \enquote{closed} term meant in CCS. In the original definition of \(\pi\)-calculus~\cite{Milner1992a, Milner1992b}, there is no notion of closed terms, as every (input) binding on a channel introduces a new and free occurrence of a variable. %
However, the language they build upon---ECCS~\cite{Engberg2000}---made this distinction clear, by separating channel constants and variables.

Once again in an attempt to mimic the \textcquote[p.~86]{Milner1993}{economy} of \(\lambda\)-calculus, but also taking inspiration from the claimed \enquote{monotheism} of the actor model~\cite{Hewitt1973}, different notions such as values, variables, or channels have been united under the common terminology of \enquote{names}.
This is at times identified as a strength, to obtain a \textcquote[p.~20]{Milner1992a}{richer calculus in which values of many kinds may be communicated, and in which value computations may be freely mixed with communications.}
However, it seems that a distinction between those notions always needs to be carefully re-introduced when discussing technically the language~\cite[p.~258, Remark 493]{Amadio2016}, extensions to it~\cite[p.~4]{Abadi2018} or possible implementations~\cites[p.~13]{Blanchet2016}{Fowler2017}.
Finally, let us note that extensions of \(\pi\)-calculus can sometimes have different binders, as \eg output binders are binding in the private \(\pi\)-calculus~\cite[p.~113]{Palamidessi2005}.

In the \(\lambda\)-calculus, being closed is what makes a term  \enquote{ready to be executed in an external environment}.
But in concurrent calculi, being a closed term---no matter how it is defined---is often not enough, as it is routine to exclude \eg terms with un-guarded operators like sum~\cite[p.~416]{Sangiorgi2001} or recursion~\cite[p.~166]{Milner1989}. %
However, these operators are sometimes not excluded from the start, even if they can never be parts of tested terms.
	The usual strategy~\cites[Remark 414]{Amadio2016}{Milner1989} is often to keep them \enquote{as long as possible}, and to exclude them only when their power cannot be tamed any more to fit the framework or prove the desired result, such as the preservation of weak bisimulation by all contexts.

In our opinion, the right distinction is not about binders of free variables, but about the role played by the syntactic objects in the theory.
As \enquote{being closed} is \begin{enumerate*}
	\item  not always well-defined, or at least changing,
	\item sometimes not the only condition
\end{enumerate*}, we would like to use the slightly more generic adjectives \emph{complete} and \emph{incomplete}--wrapped or not, in our Java terminology.
Process algebras generally study terms by \begin{enumerate*}%
\item completing them if needed, 
\item inserting them in an environment,
\item executing them,
\item observing them thanks to predicates on the execution (\enquote{terminates}, \enquote{emitted the barb \(a\)}, \etc),
\end{enumerate*}
hence constructing equivalences, preorders or metrics~\cite{Horita1997} on them.
Often, the environment is essentially made of another process composed in parallel with the one studied, %
and tweaked to improve the likeliness of observing a particular behavior: hence, we would like to think of them as tests that the observed systems has to pass, justifying %
 the terminology we will be using.
\begin{description}[leftmargin=.4em]
	\item[Processes] are \enquote{partial} programs, still under development; sometimes called \enquote{open terms}, they correspond to \emph{incomplete terms}.
	      They would be called code fragments, or snippets, in standard programming.
	\item[Systems] are \enquote{configured processes}, ready to be executed in any external environment: sometimes called \enquote{closed terms}, they correspond to \emph{complete terms}.
	      They would be functions shipped with a library in standard programming, and ready to be executed.
	\item[Tests] are defined using contexts and observations, and aims at executing and testing systems.
	      They would be \mintinline{java}{main} methods calling a library or an API in standard programming, along with a set of observables.
    \end{description}

Our terminology is close to the one used \eg in \textsc{aDpi}~\cite[Chapter 5]{Hennessy2007} or mobile ambients~\cite[Table~1]{Zappa2005}, which distinguish processes and systems.
In the literature of process algebra, the term \enquote{process} is commonly used to denote these three layers, possibly generating confusion.
We believe this usage comes from a strong desire to keep the three layers uniform, using the same name, operators and rules, but this principle is actually constantly dented (as discussed in \autoref{sec:purposes}), for reasons we expose below.
\section{Designing Layered Concurrent Languages}
\label{sec:acknowledging}

Concurrent languages could benefit from this organization %
from their conception:

\begin{description}[leftmargin=.4em]
	\item[Define processes]
	The first step is to select a set of operators called \emph{construction operators}, used by the programmer to write processes.
	Those operators should be expressive, easy to combine, with constraints as light as possible, and selected with the situation that is being modeled in mind---and not depending on whenever they fare well with not-yet-defined relations, as it is often done to privilege the guarded sum over the internal choice.
	To ease their usage, a \enquote{meta-syntax} can be used, something that is generally represented by the structural equivalence. %
(Another interesting approach is proposed in \textcquote[p.~45]{Accattoli2013}{the \(\pi\)-calculus, at a distance}, that bypasses the need for a structural equivalence without losing the flexibility it usually provides.)

	\item[Define deployment criteria] How a process can become a system ready to be executed and tested should then be defined as a series of conditions on the binding of variables, the presence or absence of some construction operators at top-level, and even the addition of \emph{deployment operators}, marking the process as ready to be deployed in an external environment\footnote{Exactly like a Java method header can use keywords---\mintinline{Java}{extends}, \mintinline{Java}{implements}, \etc---that cannot be used in a method body.
	}.
	Having a set of deployment operators that restricts%
, expands or intersects with the set of construction operators is perfectly acceptable, and it should enable the transformation of processes into systems and their composition.
	\item[Define tests]
	The last step requires to define
	\begin{enumerate*}
		\item a set of \emph{testing operators},
		\item a notion of environment constructed from those operators, along with instructions on how to place a system in it, \item a system of reduction rules regimenting how a system can execute in an environment,
		\item a set of observables, \ie a function from systems in environments to a subset of a set of atomic proposition (like \enquote{emits barb \(a\)}, \enquote{terminates}, \enquote{contains recursion operator}, \etc).
	\end{enumerate*}
\end{description}

Observe that each step uses its own set of operators and generates its own notion of context---to construct, to deploy, or to test.
Tests would be key in defining
notions of congruence, that would likely be reduction-closed, observational contextually-closed relations.
Determining if tests should wrap processes into systems or if that should be done ahead of the test itself
resonates with a long-standing debate in process algebra, and is discussed in \autoref{sec:when}.
Note that compared to how concurrent languages are generally designed, our approach is refined along two axis:
\begin{enumerate*}
	\item every step previously exposed allows the introduction of novel operators,
	\item multiple notions of systems or tests can and should co-exist in the same process algebra. %
\end{enumerate*}

\section{Addressing Existing Issues}
\label{sec:solution}

In the process algebras literature, processes and systems often have the same structure as tests and all use the same operators and contexts, to preserve and nurture a supposedly required simplicity---at least on the surface of it.
But actually, we believe that the distinction we offer is constantly used \enquote{under the hood}, without always a clear discussion, but that it %
captures and clarifies some of the choices, debates, improvements and explanations that have been proposed.%
\begin{description}[leftmargin=.4em]
	\item[Co-defining observations and contexts] Originally, the barb was a predicate~\cite[p.~690]{Milner1992}, whose definition was purely syntactic.
	      Probably inspired by the notion of observer for testing equivalences~\cite[p.~91]{Nicola1984}, an alternative definition was made in terms of parallel composition with a tester process~\cite[p.~10, Definition 2.1.3]{Madiot2015}.
	      This illustrates perfectly how the set of observables and the notion of context are inter-dependent, and that tests should always come with a definition of observable \emph{and} a notion of context: we believe our proposal could help in clarifying the interplay between observations and contexts. %
	      	One could even imagine having a series of \enquote{contexts and observations lemmas} illustrating how certain observations can be simulated by some operators, or reciprocally.%
	\item[Justifying the \enquote{silent} transition's treatment] It is routine to define relations (often called \enquote{weak}) that ignore silent (\aka \(\tau\)-) transitions, seen as \enquote{internal}.
	      This sort of transitions was dubbed \textcquote[p.~6]{Hennessy2007}{unobservable internal activity} and sometimes opposed to \textcquote[p.~230]{sangiorgi2011}{externally observable actions}.
	      While we agree that \textcquote[p.~3]{Milner1989}{\textins{t}his abstraction from internal differences is essential for any tractable theory of processes}, we would also like to stress that both can and should be accommodated, and that \enquote{internal} transition should be treated as invisible \emph{to the user}, but should still be accessible \emph{to the programmer} when they are running their own tests.

	      The question \textcquote[p.~6]{Bergstra2001}{to what extent should one identify processes differing only in their internal or silent actions?} is sometimes asked, and discussed as if it was a property of the process algebra %
		  and not something that can be \emph{internally} tuned when needed. We argue that the answer to that question is \enquote{\emph{it depends who is asking!}}: from a user perspective, internal actions should \emph{not} be observed, but it makes sense to let a programmer observe them when testing to help in deciding which process to prefer based on information not available to users. %

	\item[Letting multiple comparisions co-exist] The discussion on \(\tau\)-transitions resonates with a long debate on which notion of behavioral relation is the most \enquote{reasonable}, and---still recently---a textbook can conclude a brief overview of this issue by \textcquote[p.~160]{sangiorgi2011}{hop\textins{ing} that \textins{they} have provided enough information to \textins{their} readers so that they can draw their own conclusions on this long-standing debate}.
	Sometimes, a similar question is phrased in terms of choosing the right level of abstraction to obtain meaningful language comparisons~\cite[Section 3]{Lanese2010a}.
	      We firmly believe that the best answer to both questions is to acknowledge that different relations and comparisons tools match different needs, and that there is no \enquote{one size fits all} answer for the needs of all the variety of testers.
	      Of course, comparing multiple relations is an interesting and needed task~\cite{Fournet2005,Glabbeek1993}, but one should also state that multiple comparison tools can and should co-exist, and such vision will be encapsulated by the division we are proposing.
	\item[Embracing a feared distinction]
	      The distinction between our notions of processes and systems is rampant in the literature, but too often feared, as if it was a parenthesis that needed to be closed to restore some supposedly required purity and uniformity of the syntax.
	      A good example is probably given by mobile ambients~\cite{Zappa2005}.
	      The authors start with a two-level syntax that distinguishes between processes and systems~\cite[p.~966]{Zappa2005}.
	      Processes have access to strictly more constructors than systems~\cite[p.~967, Table~1]{Zappa2005}, that are supposed to hide the threads of computation~\cite[p.~965]{Zappa2005}.
	      A notion of \emph{system context} is then introduced---as a restriction of arbitrary contexts---and discussed, and two different ways for relations to be preserved by context are defined~\cite[p.~969, Definiton~2.2]{Zappa2005}.

	      The authors even extend further the syntax for processes with a special \(\circ\) operator~\cite[p.~971, Definition 3.1]{Zappa2005}, and note that the equivalences studied will not consider this additional constructor: we can see at work the distinction we sketched, where operators are added and removed based on different needs, and where the language needs not to be monolithic.
	      The authors furthermore introduce two different reduction barbed congruences~\cite[p.~969, Definition~2.4]{Zappa2005}---one for systems, and one for processes, with different notions of contexts---but later on prove that they coincide on systems~\cite[p~989, Theorem 6.10]{Zappa2005}.
	      It seems to us that the distinction between processes and systems was essentially introduced for technical reasons, but that re-unifying the syntax---or at least prove that systems do not do more than processes---was a clear goal right from the start.
	      We believe it would have been fruitful to embrace this distinction in a framework similar to the one we sketched: while retaining the interesting results already proven, maintaining this two-level syntax would allow to make a clearer distinction between the user's and the programmer's roles and interests, and assert that, sometimes, systems can and \emph{should} do more than processes--for instance, interacting with users!--, and can be compared using different tools.
	\item[Keeping on extending contexts]
	      We are not the first to argue that constructors can and should be added to calculi to access better discriminatory power, but without necessarily changing the \enquote{original} language.
	      The mismatch operator, for instance, has a similar feeling: \enquote{reasonable} testing equivalences~\cite[p.~280]{Boreale1995} require it, and multiple languages~\cite[p.~24]{Abadi1999} use it to provide finer-grained equivalences.
	      For technical reasons~\cite[p.~13]{Sangiorgi2001}, this operator is generally not part of the \enquote{core} of \(\pi\)-calculus, but resurfaces \emph{by need} to obtain better equivalences: we defend a liberal use of this fruitful technics, by making a clear separation between the construction operators---added for their expressivity---and the testing operators---that improve the testing capacities.

	\item[Treating extensions as different completions] It would benefit their study and usage to consider different extensions of processes algebras as different completion strategies for the same construction operators.
	      For instance, reversible~\cite{Lanese2019} or timed~\cite{Yi1991} extensions of CCS could be seen as two completion strategies---different conditions for a process to become a system---for the same class of processes, inspired from the usual CCS syntax~\cite[Chapter 28.1]{Amadio2016}.
	      Those completion strategies would be suited for different needs, as one could \eg complete a CSS process as a RCCS~\cite{Danos2004} system to test for relations such as hereditary history-preserving bisimulation~\cite{Aubert2020b}, and then complete it with time markers as a safety-critical system.
	      This would correspond to having multiple compilation, or deployment, strategies, based on the need, similar to \enquote{debug} and \enquote{real-time}, versions of the same piece of software.%
We think also of Debian's \href{https://wiki.debian.org/DebugPackage}{DebugPackage}, enabling generation of stack traces for any package, or of the \href{https://rt.wiki.kernel.org/index.php/CONFIG_PREEMPT_RT_Patch}{\texttt{CONFIG\_PREEMPT\_RT}} patch that converts a kernel into a real-time micro-kernel: both uses the same source code as their \enquote{casual} versions.

	\item[Obtaining fine-grained typing systems] The development of typing systems for concurrent programming languages is a notoriously difficult topic.
	      Some results in \(\pi\)-calculus have been solidified~\cite[Part III]{Sangiorgi2001}, but diverse difficulties remain.
	      Among them, the co-existence of multiple systems for \eg session types~\cite{denHeuvel2020}, the difficulty to tie them precisely to other type systems as Linear Logic~\cite{Caires2016}, and the doubts about the adaptation of the \enquote{proof-as-program} paradigm in a concurrent setting~\cite{Beffara2012}, make this problem active and diverse.
	      The ultimate goal seems to find a typing system that would accommodate different uses and scenarios that are not necessarily comparable.

	      Using our proposal, one could imagine easing this process by developing two different typing systems, one aimed at programmers---to track bugs and produce meaningful error messages---and one aimed at users---to track security leaks or perform user-input validation.
	      Once again, having a system developed along the layers we recommend would allow to have \eg a type system for processes only, and to erase the information when completing the process, so that the typing discipline would be enforced only when the program is being developed, but not executed.
	      This is similar to arrays of parameterized types in Java~\cite[pp.~253--258]{Leuck2013}, that checks the typing discipline at compilation time, but not at run-time. \end{description}

While this series of examples and references illustrates how our proposal could clarify pre-existing distinctions, we would like to stress that \begin{enumerate*}
	\item nothing prevents from collapsing our distinction when it is not needed,
	\item additional progresses could be made using it, as we sketch in the next section.
\end{enumerate*}

\section{Exploiting Context Awareness}
\label{sec:profit}

We would like to sketch below some possible exploitations of our frame that we believe could benefit the study and expressivity of some popular concurrent languages.

\begin{description}[leftmargin=.4em]
	\item[For CCS,]
	      we sketch below two possible improvements, the second being related to security.
	      \begin{description}[leftmargin=.4em]
		      \item[Testing for auto-concurrency]
		            Auto-concurrency (\aka auto-parallelism) is when a system has two different transitions---leading to different states---labeled with the same action~\cite[p.~391, Definition~5]{Nielsen1994}.
		            Systems with auto-concurrency are sometimes excluded as non-valid terms~\cite[p.~155]{Nicola1990} or simply not considered in particular models~\cite[p.~531]{Nielsen1988}, as the definition of bisimulation is problematic for them.
		            
		            \vspace{\baselineskip}
		            \begin{minipage}[b]{.45\textwidth}
		            Consider \eg the labeled configuration structures (\aka stable family~\cite[Section 3.1]{Winskel2017}) %
		            on the right, where the label of the event executed is on the edge and configurations are represented with \(\circ\).
		            Non-interleaving models of concurrency~\cite{Sassone1996} distinguishes between them%
		            , as \enquote{true concurrency models} would.
		        \end{minipage}
	            \begin{minipage}[b]{.50\textwidth}
	            	
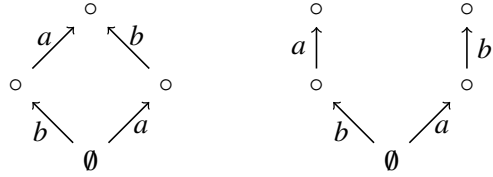
\centering
	            	\begin{tikzpicture}
	            	\node (r) at (0,0) {\(\emptyset\)};
	            	\node (g) at (1,1) {\(\circ\)};
	            	\node (d) at (-1,1) {\(\circ\)};
	            	\node (h) at (0,2) {\(\circ\)};
	            	\draw[->] (r) -- node[pos=0.8, below]{\(a\)}  (g);
	            	\draw[->] (r) -- node[pos=0.8, below]{\(b\)} (d);
	            	\draw[->] (d) -- node[pos=0.3, above]{\(a\)}  (h);
	            	\draw[->] (g) -- node[pos=0.3, above]{\(b\)}  (h);
	            	\begin{scope}[xshift=4cm]
	            	\node (s) at (0,0) {\(\emptyset\)};
	            	\node (g) at (1,1) {\(\circ\)};
	            	\node (d) at (-1,1) {\(\circ\)};
	            	\node (h1) at (-1,2) {\(\circ\)};
	            	\node (h2) at (1,2) {\(\circ\)};
	            	\draw[->] (s) -- node[pos=0.8, below]{\(a\)}  (g);
	            	\draw[->] (s) -- node[pos=0.8, below]{\(b\)} (d);
	            	\draw[->] (d) -- node[left]{\(a\)}  (h1);
	            	\draw[->] (g) -- node[right]{\(b\)}  (h2);
	            	\end{scope}
	            	\end{tikzpicture}
	            \end{minipage}
            
		            Some forms of \enquote{back-and-forth-bisimulations} cannot discriminate between them %
		            if \(a = b\)~\cite{Phillips2007}.
		            While not being able to distinguish between those two terms may make sense from an external---user's---point of view, we argue that a programmer should have access to an internal mechanism that could answer the question \enquote{\emph{Can this process perform two barbs with the same label at the same time?}}.
		            Such an observation---possibly coupled with a testing operator---would allow to distinguish between \eg \(!a.P \mid !a.P\) and \(!a.P\), that are generally taken to be bisimilar, and would re-integrate auto-concurrent systems---that are, after all, unjustifiably excluded---in the realm of comparable systems.

		      \item[Representing man-in-the-middle] One could add to the testing operators an operator \(\nabla a.P\), which would forbid \(P\) to act silently on channel \(a\).
		            This novel operator would add the possibility for the environment to \enquote{spy} on a determined channel, as if the environment was controlling (part of) the router of the tested system.
		            One could then reduce \enquote{normally} in a context \(\nabla a [\square]\) if the channel is still secure:
		            \begin{align}
			            \nabla a ( b.Q\mid \bar{b}.P) & \to^{\tau} \nabla a (Q\mid  P) \tag{If \(a\neq b\)}
		            \end{align}
		            But in the case where \(a = b\), the environment could intercept the communication and then decide to forward, prevent, or alter it.
		            Adding this operator to the set of testing operators would for instance open up the possibility of interpreting \(\restr{a} P\) as an operation securing the channel \(a\) in \(P\), enabling the study of relations \(\sim\) that could include \eg
		            \begin{align}
			            \nabla a (\restr{a}{P|Q}  & \sim \nabla a (\restr{b}{P[a/b]|Q[a/b]}) \tag{For \(b\) not in the free names of \(P\) nor \(Q\)} \\
			            \restr{a}{\nabla a (P|Q)} & \sim \nabla a ( P|Q) \tag{Uselessness of securing a hacked channel}
		            \end{align}
		            While the first rule enforces that, once secured, channel names are \(\alpha\)-equivalent (the process can decide to migrate to a different channel without being spied on), the second illustrates that, once a channel is tapped, a process cannot obtain any confidentiality on it anymore.

	      \end{description}

	\item[In \(\pi\)-calculus,]
	      all tests are instantiating contexts (in the sense that the term tested needs to be either already closed, or to be closed by the context), and all instantiating contexts use only construction operators, and hence are \enquote{construction contexts}.
	      This situation corresponds to Situation A in \autoref{fig:pi-contexts}.

	      We believe the picture could be much more general, with tests having access to \emph{more constructors}, and not needing to be instantiating---in the sense that completion can be different from closedness---, so that we would obtain Situation B in \autoref{fig:pi-contexts}.
	      While we believe this remark applies to most of the process algebras we have discussed so far, it is particularly salient in \(\pi\)-calculus, where the match and mismatch operators have been used \textcquote[p.~57]{FuY2003a}{to internalize a lot of meta theory}, standing \enquote{inside} the \enquote{Construction operators} circle while most authors seem to agree that they would prefer not to add it to the internals of the language\footnote{To be more precise: while \textcquote[p.~526]{Parrow1993}{most occurences of matching can be encoded by parallel composition \textins{…,} mismatching cannot be encoded in the original \(\pi\)-calculus}, which makes it somehow suspicious.}.
	      It should also be noted that the mismatch operator---in its \enquote{intuitionistic} version---furthermore \enquote{tried to escape the realm of instantiating contexts} by being tightly connected~\cite{Horne2018} to \emph{quasi-open bissimilarities}~\cite[p.~300, Definition 6]{Sangiorgi2001c}, which is a subtle variation on how substitutions can be applied by context to the terms being tested.

	      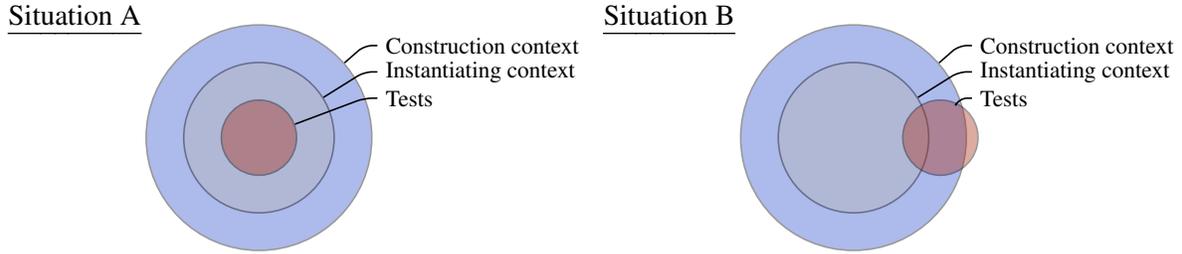
\begin{figure}
	      	{\centering
	      		
		      \begin{tikzpicture}[scale=0.35]
			      \node at (-2, 4.5) {\ul{Situation A}};
			      \begin{scope}[xshift=5cm]
				      \drawsituation{0}{0.8}
			      \end{scope}
		      \end{tikzpicture}
		      \begin{tikzpicture}[scale=0.35]
			      \node at (-2, 4.5) {\ul{Situation B}};
			      \begin{scope}[xshift=5cm]
				      \drawsituation{3.3}{0.8}
			      \end{scope}
		      \end{tikzpicture}
		    
	    	}
		      \caption{Opening up the testing capacities of \(\pi\)-calculus}
		      \label{fig:pi-contexts}
	      \end{figure}

	      Having a notion of \enquote{being complete} not requiring closedness could be useful when representing distributed programming, where \textcquote[p.~250]{Hashimoto2001}{one often wants to send a piece of code to a remote site and execute it there. \textins{…} \textins{T}his feature will greatly enhance the expressive power of distributed programming\textins{ by } send\textins{ing} an open term and to make the necessary binding at the remote site.}
	      We believe that maintaining the possibility of testing \enquote{partially closed}---but still complete---terms would enable a more theoretical understanding of distributed programming and remote compilation.

	\item[Distributed \(\pi\)-calculus,] could explore the possible differences between two parallelisms: between threads in the same process---in the Unix sense---and between units of computation.
	      Such a distinction could be rephrased thanks to two parallel operators, one on processes and the other on systems.
	      Such a distinction would allow to observationally distinguish \eg the execution of a program with two threads on a dual-core computer and the execution of two single thread programs on two single-core computers.
	\item[For cryptographic protocols,]
	      we could imagine representing encryption of data as a special context \(\mathcal{E}[\square]\) that would transform a process \(P\) into an encrypted system \(\mathcal{E}[P]\), and make it un-executable unless \enquote{plugged} in an environment \(\mathcal{D}[\square]\) that could decrypt it.
	      This could allow the applied \(\pi\)-calculus~\cite{Abadi2018} to become more expressive and to be treated as a decoration of the pure \(\pi\)-calculus more effectively.
	      This could also, as the authors wish, make \textcquote[p.~35]{Abadi2018}{the formalization of attackers as contexts \textins{…} continue to play a role in the analysis of security protocols}.
	      
	      Recent progresses in the field of verification of cryptographic protocols~\cite{Baelde2021} hinted in this direction as well.
	      By taking \textcquote[p.~12]{Baelde2021}{\textins{t}he notion of test \textins{to}  be relative to an environment}, a formal development involving \enquote{frames}~\cite[Definition 2.3]{Baelde2021} can emerge and give flesh to some ideas expressed in our proposal.
	      It should be noted that this work also \textcquote[p.~12]{Baelde2021}{enrich\textins{…} processes with a success construct}, that cannot be used to construct processes, to construct \enquote{experiments}.
\end{description}

\section{Concluding Remarks}
We conclude by discussing related approaches, by casting a new light on a technical issue related to barbed congruences, by offering the context lemmas a new interpretation, and by coming back to our motivations.

\subsection{An Approved and Promising Perspective}
\label{ssec:approved-persp}

We would like to stress that our proposal resonates with previous comments, and should not be treated as an isolated historical perspective that will have no impact on the future.

In the study of process algebras, in addition to the numerous hints toward our formalism that we already discussed, there are at least two instances when the power of the \enquote{testing suite} was explicitly discussed~\cite[Remark 5.2.21]{Sangiorgi2001b}.
In a 1981 article, it is assumed that \textcquote[p.~32]{Milner1981}{by varying the ambiant (\enquote{weather}) conditions, an experimenter} can observe and discriminate better than a simple user could.
Originally, this idea seemed to encapsulate two orthogonal dimensions: the first was that the tester could run the tested system any number of times, something that would now be represented by the addition of the replication operator \(!\) to the set of testing operators.
The second was that the tester could enumerate all possible non-deterministic transitions of the tested system.
This second dimension gave birth to \textcquote[p.~1]{Larsen1991}{a language for testing concurrent processes} that is more powerful than the language used to write the programs being tested.
In this particular example, the tester has access to a termination operator and probabilistic features that are not available to the programmer: as a result, the authors \textcquote[p.~19]{Larsen1991}{may distinguish non-bisimilar processes through testing}.

Looking forward, the vibrant field of secure compilation made a clear-cut distinction between \enquote{target language contexts} representing adversarial code and programmers' \enquote{source context} to explore property preservation of programs~\cite{Abate2019}.
This perspective was already partially at play in the spi calculus for cryptographic protocols~\cite[p.~1]{Abadi1999}, where the attacker is represented as the \enquote{environment of a protocol}.
We believe that both approaches---coming from the secure compilation, from the concurrency community, but also from other fields---concur to the same observation that the environment---formally captured by a particular notion of context---deserves an explicit and technical study to model different interactions with systems, and need to be detached from \enquote{construction} contexts.

\subsection{When Should Contexts Come into Play?}
\label{sec:when}
The interesting question of \emph{when} to use contexts when testing terms~\cite[pp.~116--117, Section 2.4.4]{Sangiorgi2001} raises a technical question that is put under a different perspective by our analysis. Essentially, the question is whether the congruences under study should being \emph{defined} as congruences (\eg reduction-closed barbed congruence~\cite[p.~116]{Sangiorgi2001}), or being defined in two steps, \ie as the contextual closure of a pre-existing relation (\eg strong barbed congruence~\cite[p.~61, Definition 2.1.17]{Sangiorgi2001}, which is the contextual closure of strong barbed bisimilarity~\cite[p.~57, Definition 2.1.7]{Sangiorgi2001})?

Indeed, bisimulations can be presented as an \textcquote{Stirling1995}{interaction game} generally played as
\begin{enumerate*}
	\item Pick an environment for both terms (\ie, complete them, then embed them in the same testing environment),
	\item Have them \enquote{play} (\ie have them try to match each other's step).
\end{enumerate*}
But a more dynamic version of the game let picking an environment \emph{be part of the game}, so that each process can not only pick the next step, \emph{but also in which environment it needs to be performed}.
This version of the game, called \textcquote{Montanari1992}{dynamic observational congruence}, provides a better software modularity and reusability, as it allows to study the similarity of terms that can be re-configured  \enquote{on the fly}.
Embedding the contexts in the definitions of the relations is a strategy that was also used to obtain behavioral characterization of theories~\cite[p.~455, Proposition 3.24]{Honda1995}, and that corresponds to open bisimilarities~\cite[p.~77, Proposition 3.12]{Sangiorgi1996}%

Those two approaches have been extensively compared and studied--still are~\cite[p.~24]{Abadi2018}---but to our knowledge they rarely co-exist, as if one had to take a side at the early stage of the language design, instead of letting the tester decide later on which approach is best suited for what they wish to observe.
We argue that both approaches are equally valid, \emph{provided we acknowledge they play different roles}.

\begin{figure*}
	{
		\centering
		\begin{tikzpicture}[
				block/.style={
						rectangle,
						thick,
						text width=7em,
						align=center,
						rounded corners,
						minimum height=2em
					}
			]

			\node at (-4, 0) {\ul{Situation A}};
			\node [block, fill=licsred!20, draw=licsred, label={[label distance=-.3em]\emph{Write}}] (write) at (0,0) {Process};
			\node [block, fill=licscomp2!20, draw=licscomp2, label={[label distance=-.3em]\emph{Complete}}] (complete) at (4,0) {System};
			\node [block, fill=licsgray!20, draw=licsgray, label={[label distance=-.3em]\emph{Compare}}] (compare) at (8, 0) {Test};
			\draw[thick,dotted,draw=licscomp2] ($(write.north west)+(-0.5,0.5)$) node[above right]{Programmer} rectangle ($(complete.south east)+(0.5,-0.5)$) ;
			\draw[thick,dotted,draw=licscomp2] ($(compare.north west)+(-0.5,0.5)$) node[above right]{Tester} rectangle ($(compare.south east)+(0.5,-0.5)$) ;
			\draw [->, thick] (write) to (complete);
			\draw [->, thick] (complete) to (compare);
			\begin{scope}[yshift=-3cm]
				\node at (-4, 0) {\ul{Situation B}};
				\node [block, fill=licsred!20, draw=licsred, label={[label distance=-.3em]\emph{Write}}] (write) at (0,0) {Process};
				\node [block, fill=licscomp2!20, draw=licscomp2, label={[label distance=-.3em]\emph{Complete}}] (complete) at (4,0) {System};
				\node [block, fill=licsgray!20, draw=licsgray, label={[label distance=-.3em]\emph{Compare}}] (compare) at (8, 0) {Test};
				\draw[thick,dotted,draw=licscomp2] ($(write.north west)+(-0.5,0.5)$) node[above right]{Programmer} rectangle ($(write.south east)+(0.5,-0.5)$) ;
				\draw[thick,dotted,draw=licscomp2] ($(complete.north west)+(-0.5,0.5)$) node[above right]{Tester} rectangle ($(compare.south east)+(0.5,-0.5)$) ;
				\draw [->, thick] (write) to (complete);
				\draw [<->, thick] (complete) to (compare);
			\end{scope}
		\end{tikzpicture}
	}
	\caption{Distinguishing between completing strategies}
	\label{fig:closing}
\end{figure*}
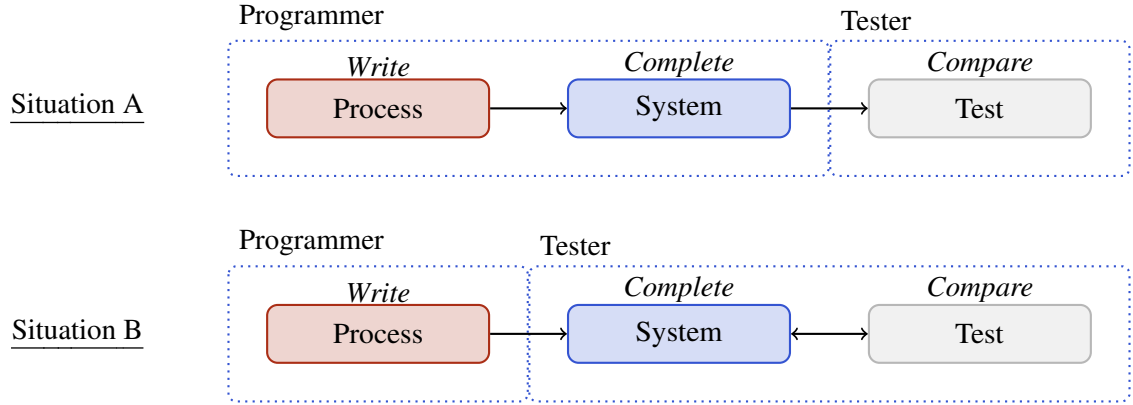

This question of \emph{when are the terms completed?} can be rephrased as \emph{what is it that you are trying to observe?}, or even \emph{who is completing them?}: is the completion provided by the programmer, once and for all, or is the tester allowed to explore different completions and to change them as the tests unfold?
Looking back at our Java example from \autoref{sec:Java-example}, this corresponds to letting the tester repeatedly tweak the parameter or return type of the wrapping from \mintinline{Java}{int} to \mintinline{Java}{long}, allowing them to have finer comparisons between snippets.
In this frame, moving from the \emph{static} definition of congruence to \emph{dynamic} one would corresponds to going from Situation A to Situation B in \autoref{fig:closing}.
This illustrates two aspects worth highlighting: \begin{enumerate}
	\item Playing on the variation \enquote{\emph{should I complete the terms before or during their comparison?}} is not simply a technical question, but reflects a choice between two different situations equally interesting. \item This choice can appeal to different notions of systems, completions and tests: for instance, while completing a process before testing it (Situation A) may indeed be needed when the environment represents an external deployment platform, it makes less sense if we think of the environment as part of the development workflow, in charge of providing feedback to the programmer or as a powerful attacker than can manipulate the conditions in which the process is executed (Situation B).
\end{enumerate}
If completion is seen as compilation, this opens up the possibility of studying how the bindings performed \emph{by the user}, on \emph{their} particular set-up, during a \emph{remote} compilation, can alter a program.
One can then compare different relations---some comparing source code's releases, some comparing binaries' releases---to get a better, fuller, picture of the program.

\subsection{Penetrating Context Lemmas' Meanings}
\label{sec:context}

What is generally refereed to as \emph{the} context lemma\footnote{At least, in process algebra, as the same name is used for a different type of meaning in \eg \(\lambda\)-calculus~\cite[p.~6]{Milner1977}.}  is actually a series of results stating that considering all the operators when constructing the context for a congruence may not be needed.
For instance, it is equivalent to define the barbed congruence~\cite[p.~95, Definition 2.4.5]{Sangiorgi2001} as the closure of barbed bisimilarity under all context, or only under contexts of the form \([\square] \sigma\mid P\) for all substitution \(\sigma\) and term \(P\).
In its first version~\cite[p.~432, Lemma 5.2.2]{Pierce1996}, this lemma had additional requirements \eg on sorting contexts, but the core idea is always the same: \enquote{\emph{there is no need to consider all contexts to determine if a relation is a congruence, you can consider only contexts of a particular form}}.

The \enquote{flip side} of the context lemma is what we would like to call the \enquote{anti-context pragmatism}: whenever a particular type of operator or context prevents a relation from being a congruence, it is tempting to simply exclude it.
For instance, contexts like \([\square] + 0\) are routinely removed---as we discussed in \autoref{sec:purposes}---to define the barbed congruence of \(\pi\)-calculus, or contexts were restricted to what is called harnesses in the mobile ambients calculus~\cite{Gordon2003} before proving such results.
As strong bisimulation~\cite[p.~514, Definition 1]{Parrow2001} is not preserved by input prefix~\cite[p.~515, Proposition 4]{Parrow2001} but is by all the other operators, it is sometimes tempting to simply remove input prefix from the set of constructors allowed at top-level in contexts, which is what non-input contexts~\cite[p.~62, Definition 2.1.22]{Sangiorgi2001} do, and then to establish a context lemma for this limited notion of context.

Taken together, those two remarks produce a strange impression: while it is mathematically elegant and interesting to prove that weaker conditions are enough to satisfy an interesting property, it seems to us that this result is sometimes \enquote{forced} into the process algebra by having ahead of time excluded all the operators that would not fit, hence producing a result that is not only weaker, but also somehow artificial, or even tautological.
Furthermore the criteria of \enquote{not adding any discriminating power} should not be a positive criterion when deciding if a context belongs to the algebra: on the opposite, one would want contexts to \emph{increase} the discriminating power---as for the mismatch operator---and not to \enquote{conform} to what substitution and parallel composition have already decided.

Context lemmas seem to embrace an uncanny perspective: instead of being used to prove properties about tests more easily, they should be considered from the perspective of the ease of use of testing systems.
Stated differently, we believe that the set of testing operators should come first, and then \emph{then}, if the language designer wishes to add operators to ease the testers' life, they can do so providing they obtain a context lemma proving that those operators do not alter the original testing capacities.
Once again, varying the testing suite is perfectly acceptable, but once fixed, \emph{the context lemma is simply present to show that adding some testing operators is innocent, that it will simply make testing certain properties easier}.

\subsection{Embracing the Diversity}
\label{sec:what}
Before daring to submit a non-technical paper, we tried to conceive a technical construction that could convey our ideas.
In particular we tried to build a syntactic (even categorical) meta-theory of processes, systems and tests.
We wanted to define congruences in this meta-theory, and to answer the following question: what could be the minimal requirements on contexts and operators to prove a generic form of context lemma for concurrent languages?

However, as the technical work unfolded, we realized that the definitions of contexts, observations, and operators, were so deeply interwoven that it was nearly impossible to extract any general or useful principle.
Context lemmas use specific features of languages, in a narrow sense, %
as for instance no context lemma can exist in the \enquote{Situation B} of \autoref{fig:closing}~\cite[p.~117]{Sangiorgi2001}, and we were not able to find a unifying framework.
This also suggests that context lemmas are often \emph{fit} for particular process algebras \emph{by chance}, and dependent intrinsically of the language considered, for no deep reasons.

This was also liberating, as all the nuances of languages we had been fighting against started to form a regular pattern: every single language we considered exhibited (at least parts of) the structure we sketched in the present proposal.
Furthermore, our framework was a good lens to read and answer some of the un-spoken questions suggested in the margin or the footnotes---but rarely upfront---of the multiple research papers, lecture notes and books we consulted.
So, even without mathematical proofs, we consider this contribution a good way of stirring the community, and to question the traditional wisdom.

It seems indeed to us that there is nothing but benefits in altering the notion of context, as it is actually routine to do so, even recently~\cite{hirschkoff2020}, and that stating the variations used will only improve the expressiveness of the testing capacities and the clarity of the exposition.

It is a common trope to observe the immense variety of process calculi, and to sometimes wish there could be a common formalism to capture them all---to this end, \emph{the} \(\pi\)-calculus is often considered the best candidate.
Acknowledging this diversity is already being one step ahead of the \(\lambda\)-calculus---that keeps forgetting that there is more than one \(\lambda\)-calculus, depending on the evaluation strategy and on features such as sharing~\cite{Accattoli2019}---and this proposal encourages to push the decomposition into smaller languages even further, as well as it encourages to see whole theories as simple \enquote{completion} of standard languages.
As we defended, breaking the monolithic status of context%
\footnote{%
	This may be a good place to mention that this monolithicity probably comes in part from the original will of making \eg CCS a programming \emph{and} specification language---an original perspective that was reminded to us by \href{http://www-sop.inria.fr/members/Ilaria.Castellani/}{Ilaria Castellani}, who we wish to thank. The specification was supposed to be the program itself, that would be easy to check for correctness: the goal was to make it \textcquote[p.~1]{Milner1986}{possible to describe existing systems, to specify and program new systems, and to argue mathematically about them, all without leaving the notational framework of the calculus}. This original research project slightly shifted---from specifying programs to specifying behaviors---but that original perspective remained.%
	}
will actually make the theory and presentation follow more closely the technical developments, and liberate from the goal of having to find \emph{the} process algebra with \emph{its unique} observation technique that would capture all possible needs.

\section{Postscript}
\label{sec:postscript}

The tone and scope of this essay may seem \enquote{timeless}, but discussions during the \emph{14th Interaction and Concurrency Experience} (particularly with \href{https://www.imm.dtu.dk/~alcsc/}{Alceste Scalas} and \href{https://cs.gssi.it/emilio.tuosto/}{Emilio Tuosto}) and with colleagues (among which \href{http://www.lsv.fr/~baelde/}{David Baelde} and \href{https://satoss.uni.lu/members/ross/}{Ross Horne}) finished anchoring it to very concrete and timely questions.
We would like to briefly sketch a handful of future themes, hoping to faithfully record the fascinating discussions that led to them.

\begin{enumerate}
	\item Semantical approaches to security revealed multiple small gaps that are hard to explain or solve with usual notions of contexts and equivalences. 
	For instance, the study of equivalence properties led to distinguish between semantics where all the communications have to be made via the environment (\ie, the attacker) and semantics where communications inside the process and communications between the process and its environment are treated differently.
	The (surprising) result~\cite{Babel2020} is that while both treatments coincide for reachability properties, they are incomparable for indistinguishability properties.
	The resulting \emph{classical}, \emph{private} and \emph{eavesdropping} semantics each yield their own may-testing and observational equivalences, that could possibly be elegantly captured by our frame.
	The interested reader should refer to a recent discussion~\cite[Section 4.1.6]{Baelde2021} for more insight on this issue.
	\item Our discussion could also be used to bridge possible gaps between process algebras and game semantics, or at least introduce a gradation between them.
	Indeed, while \enquote{usual} algebraic contexts are extremely passive and with limited (static) knowledge of the term placed inside of them, strategies generally have all laterality to evolve, and to store information about the play unfolding.
	We would like to argue, following a classical paper~\cite[Section 5]{Laird2005}, that maybe processes (or, in our case, contexts) could represent strategies, and that there could be a gradation in the \enquote{dynamicity} of contexts: from very statical ones to more dynamical and reactive ones.
	This could be used \eg to study the open barbed bisimilarity between processes with free variables, by allowing contexts to use and retain information about permitted submissions.
	\item Reversible CCS (RCCS)~\cite{Danos2004} and CCS with keys (CCSK)~\cite{Phillips2006} are two extensions to CCS aiming at formalizing reversible concurrent computation.
	They have been compared and studied, but, as it turns out, they actually are the two faces of the same coin~\cite{Lanese2019}.
	However, when defining contexts, and despite the similarities in the definitions, two recent studies~\cite{Aubert2021d, Lanese2021} came to exactly opposite conclusions: while the RCCS-inspired system seems to provide definitive evidence that \emph{no relation can be a congruence if the context can change the history of the process}~\cite[Theorem 2]{Aubert2021d}, the CCSK approach came to the conclusion that a particular barbed bisimilarity was a congruence~\cite[Corollary 5.12]{Lanese2021}. Does the difference rest only on the definition of context, or is there a more subtle distinction at work? It is our hope that the approach sketched here could help to solve this mystery.
\end{enumerate}

Last but not least, an \enquote{easy} exercise to illustrate the relevance of our perspective would be to study a Morris-style equivalence of the programming language Java, using the convenient, \enquote{simplified}, Featherweight Java~\cite{Igarashi2001}.
It is our hope that this lead, on which we will be working in the near future, will fruitfully illustrate the benefits of the distinction we detailed in this paper.

\begin{acknowledgements}
	The author wish to thank the organizer of \href{https://www.discotec.org/2021/ice}{ICE 2021} for organizing this welcoming, open workshop, as well as the reviewers who kindly shared their comments, suggestions and insights with us. This paper benefited a lot from them, as well as from the discussions mentioned in \autoref{sec:postscript}.
\end{acknowledgements}

\newpage
\bibliographystyle{eptcs}
\bibliography{bib/bib}
\end{document}